\documentclass[a4paper]{spie}  

\usepackage[utf8]{inputenc}
\usepackage[T1]{fontenc}
\usepackage[english]{babel}
\usepackage{amsfonts}
\usepackage{amsmath}
\usepackage{graphicx}
\usepackage{color}
\usepackage[colorlinks=true, allcolors=blue]{hyperref}

\newcommand{\widfig}[2]{\includegraphics[width=#1\columnwidth]{#2}}

\usepackage{esvect}
\renewcommand{\vec}[1]{\vv{#1}}

\newcommand{\mii}{\mathrm{i}}
\newcommand{\mee}{\mathrm{e}}

\newcommand{\mat}[1]{\mbox{\boldmath{$\mathrm{#1}$}}}

\newcommand{\identmat}{\mat I}
\newcommand{\diagmat}{\mathop{\mathrm{diag}}}
\newcommand{\transp}{\mathrm{T}}

\newcommand{\submat}[3]{\left\{#1\right\}_{#2}^{#3}}
\newcommand{\transv}{\perp}

\title{Cluster synchronization of starlike networks with normalized
  Laplacian coupling: master stability function approach}

\author[a]{Pavel V. Kuptsov}%
\author[a]{Anna V. Kuptsova}%

\affil[a]{Institute of electronics and mechanical engineering, Yuri
  Gagarin State Technical University of Saratov, Politekhnicheskaya
  77, Saratov 410054, Russia}%

\authorinfo{Send correspondence to P.V.K. E-mail:
  p.kuptsov@rambler.ru}

\begin{document}

\maketitle

\begin{abstract}
  A generalized model of starlike network is suggested that takes into
  account non-additive coupling and nonlinear transformation of
  coupling variables. For this model a method of analysis of
  synchronized cluster stability is developed. Using this method three
  starlike networks based on Ikeda, predator-prey and H\'enon maps are
  studied.
\end{abstract}

\keywords{starlike network, master stability function, cluster
  synchronization, full chaotic synchronization}

\section{Introduction} 

In this paper we consider starlike networks that consist of a single
hub node connected with all other nodes, and subordinate nodes that
have only one connection with the hub and are not connected with each
other.

The starlike networks is an interesting object of study due to the
following reasons. First, they are the simplest and regular
representatives of so called scale-free networks that attract a lot of
interest now as models for a large variety of natural
systems\cite{Boccaletti2006175,CxNetwTopDynSyn2002}. These networks
are ``scale-free'' since their node degree distributions have power
law shapes. As a result a small number of nodes hold a major balk of
links while the rest of nodes have few
connections~\cite{ScaleFreeNetw}. Another reason for interest to
starlike networks is that their dynamics is amazingly rich. The main
their feature is multistability when the number of attractors is very
high, and their basins have fractal boundaries that are highly
interwoven. It is known that in this case the dynamics is very
sensible to the initial state: even tiny perturbation results in
arriving at new regime. Moreover, the ranges of existence of
particular attractors can be narrow so that the qualitative behavior
of the system can change dramatically when its parameters are slightly
varied~\cite{Feudel2008,Pisarchik2014167,WildStars}.

In the present paper we suggest a generalized model of starlike
network taking into account non-additive coupling and nonlinear
transformation of coupling variables. For this model we develop an
analysis of stability of synchronized clusters generalizing the idea
of master stability function~\cite{MSF98}. Using this method we study
the dynamics of three starlike networks based on Ikeda, predator-prey
and H\'enon maps.

\section{Generic model}\label{sec:generic}

We are going to consider dynamical networks of $N$ elements with nodes
occupied by $M$ dimensional identical discrete time systems:
\begin{gather}
  \label{eq:netw_gen}
  x_n^{(m)}(t+1)=f^{(m)}[\vec x_n(t),h_n^{(m)}(t)],\\
  \label{eq:netw_gen_h}
  h_n^{(m)}(t)=\epsilon_m\sum_{j=1}^N\ell_{nj}g^{(m)}[\vec x_j(t)].
\end{gather}
Here upper indexes $m=1,\ldots,M$ enumerate local variables of maps,
and lower ones $n=1,\ldots,N$ runs along network nodes. Functions
$f^{(m)}(\vec x,h)$ determine a node map, where vector
$\vec x=(x^{(1)},\ldots,x^{(M)})^\transp$ is a shorthand for its $M$
arguments, and $h$ is its $(M+1)$th argument responsible for the
coupling. Function $g^{(m)}(\vec x)$ depends on $M$ arguments and
defines nonlinear transformation of coupling variables. The network
structure is given by a $N\times N$ matrix $\mat L=\{\ell_{nj}\}$. Its
particular structure will be specified below. Finally, the coupling
strength is controlled by $\epsilon_m$.

Any stability analysis requires Jacobian matrix and corresponding
equation for infinitesimal perturbations to a
trajectory. Differentiating $x_n^{(m)}(t+1)$ by $x_k^{(i)}(t)$ one
obtains:
\begin{equation}
  \label{eq:netw_gen_variat}
    \delta x_n^{(m)}(t+1)=\sum_{i=1}^M\sum_{k=1}^N
    \big(
    \delta_{nk}f^{(m)}_{in}+f^{(m)}_{M+1,n}\epsilon_m\ell_{nk}g^{(m)}_{ik}
    \big)
    \delta x_k^{(i)}(t)
\end{equation}
where $\delta x_n^{(m)}(t)$ is a perturbation to $m$th variable of
$n$th node, and
\begin{equation}
  f^{(m)}_{in}\equiv \frac{\partial}{\partial x^{(i)}}f^{(m)}
  (\vec x_n,h_n^{(m)}),\;\;
  f^{(m)}_{M+1,n}\equiv \frac{\partial}{\partial h}f^{(m)}(\vec x_n,h_n^{(m)}),\;\;
  g^{(m)}_{ik}\equiv \frac{\partial}{\partial x^{(i)}}g^{(m)}(\vec x_k).
\end{equation}
Grouping perturbations to variables with identical indexes into $N$
dimensional vectors
$\vec{\delta x}^{(m)}=(\delta x_1^{(m)},\ldots,\delta
x_N^{(m)})^\transp$ one can rewrite Eq.~\eqref{eq:netw_gen_variat} as
\begin{equation}
  \label{eq:netw_block_variat}
  \vec{\delta x}^{(m)}(t+1)=\sum_{i=1}^M\mat B_{mi}\,\vec{\delta x}^{(i)}(t)
\end{equation}
where 
\begin{equation}
  \label{eq:jac_cell}
  \mat B_{mi}=\mat F_{mi}+\epsilon_m\mat \Phi_m\mat L\mat G_{mi}
\end{equation}
and
\begin{equation}
  \label{eq:jac_f_ph_g}
  \begin{aligned}
    \mat F_{mi}&=\diagmat\{f^{(m)}_{in},n=1,\ldots,N\}\\
    \mat \Phi_{m}&=\diagmat\{f^{(m)}_{M+1,n},n=1,\ldots,N\}\\
    \mat G_{mi}&=\diagmat\{g^{(m)}_{in},n=1,\ldots,N\}
  \end{aligned}
\end{equation}
Thus, the Jacobian matrix of the network~\eqref{eq:netw_gen} is
$M\times M$ block matrix whose cells are $N\times N$ matrices
$\mat B_{mi}$

\section{Starlike networks}\label{sec:stars}

\begin{figure}
  \begin{center}
    \widfig{0.25}{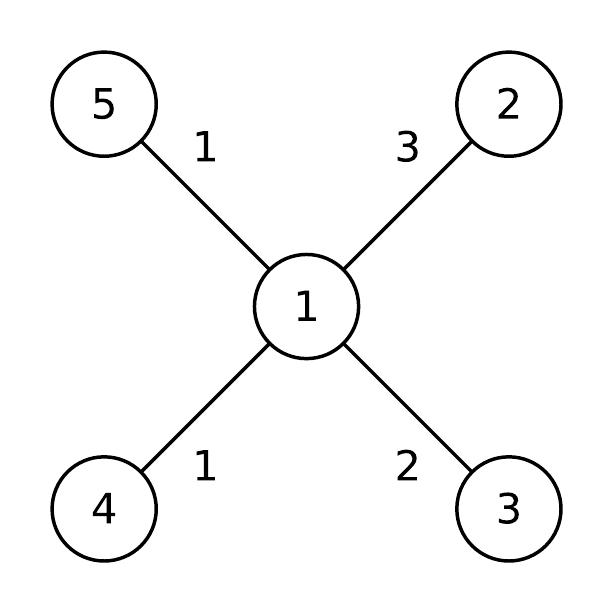}
  \end{center}
  \caption{\label{fig:star5}Starlike network with $N=5$ nodes. Links
    have positive integer weights.}
\end{figure}

Starlike networks that we consider in this paper has a single hub node
and some number of subordinate ones. Each subordinate is connected
with the hub and no subordinates are connected with each other. Each
connection can have integer positive weight. An example of the star
with $N=5$ nodes is shown in Fig.~\ref{fig:star5}. The weights of
links will be denoted as $w_i$, where $i=2,\ldots, N$ is an index of
the corresponding subordinate node. In Fig.~\ref{fig:star5} $w_2=3$,
$w_3=2$, $w_4=1$, and $w_5=1$.

The adjacency matrix of a starlike network can be written as
\begin{equation}
  \label{eq:star_gen}
  \mat A=
  \left(
  \begin{array}{ll}
    0 & \mat w \\
    \mat w^\transp & \submat{\mat 0}{2}{N}
  \end{array}
  \right),
\end{equation}
where $\mat w=(w_2,w_3,\ldots,w_{N})$ and $\submat{\mat 0}{2}{N}$ is
$N-1\times N-1$ matrix of zeros. Let $\mat K$ be a diagonal matrix of
row sums of $\mat A$. The normalized Laplacian coupling matrix reads:
\begin{equation}
  \label{eq:lapl_mat}
  \mat L=\mat K^{-1}\mat A-\identmat.
\end{equation}
We already considered this type of coupling matrices for scale-free
and starlike networks of H\'enon maps\cite{NWL2014,WildStars}.

Substituting this $\mat L$ to equation for a cell $\mat B_{mi}$ of the
Jacobian matrix~\eqref{eq:jac_cell} one gets:
\begin{equation}
  \label{eq:star_cell}
  \mat B_{mi}=
  \left(
    \begin{array}{ll}
      \submat{\mat F_{mi}}{1}{1}-\epsilon_m\submat{\mat G_{mi}}{1}{1} 
        \submat{\mat \Phi_{mi}}{1}{1} &
      \epsilon_m\submat{\mat \Phi_{mi}}{1}{1} \mat w 
        \submat{\mat G_{mi}}{2}{N}/W \\
      \epsilon_m\submat{\mat G_{mi}}{1}{1} 
        \submat{\mat \Phi_{mi}}{2}{N}\mat d^\transp  &
      \submat{\mat F_{mi}}{2}{N}-\epsilon_m\submat{\mat \Phi_{mi}}{2}{N}
        \submat{\mat G_{mi}}{2}{N}
    \end{array}
  \right),
\end{equation}
where $\mat d=\left(1,1,\ldots,1\right)$, and
$W=\sum_{i=2}^{N}w_i$. Symbol $\submat{\mat F}{i}{j}$ stands for
square diagonal submatrix including elements form $i$th to $j$th of
the diagonal matrix $\mat F$, and the same for other matrices.

\section{Cluster synchronization and hierarchy of invariant
  manifolds}\label{sec:manif}

The normalized Laplacian coupling matrix~\eqref{eq:lapl_mat} admits the full
synchronization of the whole star when
$\vec x_1(t)=\cdots=\vec x_N(t)$ for any $t$. Moreover the full
synchronization of any number of subordinate nodes is also possible,
i.e., cluster synchronization can be observed. To understand why this
is the case let us write $h_n^{m}$, see Eq.~\eqref{eq:netw_gen_h}, for
a starlike network explicitly:
\begin{align}
  \label{eq:netw_h_star}
  h_1^{(m)}&=\epsilon_m\left[
    \sum_{j=2}^N w_jg^{(m)}(\vec x_j)/W-g^{(m)}(\vec x_1)
  \right],\\
  \label{eq:netw_h_star2}
  h_n^{(m)}&=\epsilon_m\left[g^{(m)}(\vec x_1)-g^{(m)}(\vec
                    x_n)\right],\;n\geq 2.
\end{align}
One can see that if nodes $n_1$ and $n_2$ have identical states at
$t=t_1$, where $n_1\geq 2$ and $n_2\geq 2$, then corresponding
$h_{n}^{(m)}$ are also identical and hence these states remain
identical at $t=t_1+1$. In the same way any number of subordinates can
keep identical states forming a synchronization cluster. Moreover
several clusters are also possible.

The existence of the clusters means that the phase space of the
considered networks involves corresponding invariant synchronization
manifolds. The manifolds can be denoted with sequences of $N$
digits. Zeros represent variables corresponding to non-synchronized
nodes, while identical non-zero digits indicate that the corresponding
node variables coincide.

\begin{figure}
  \begin{center}
    \widfig{0.8}{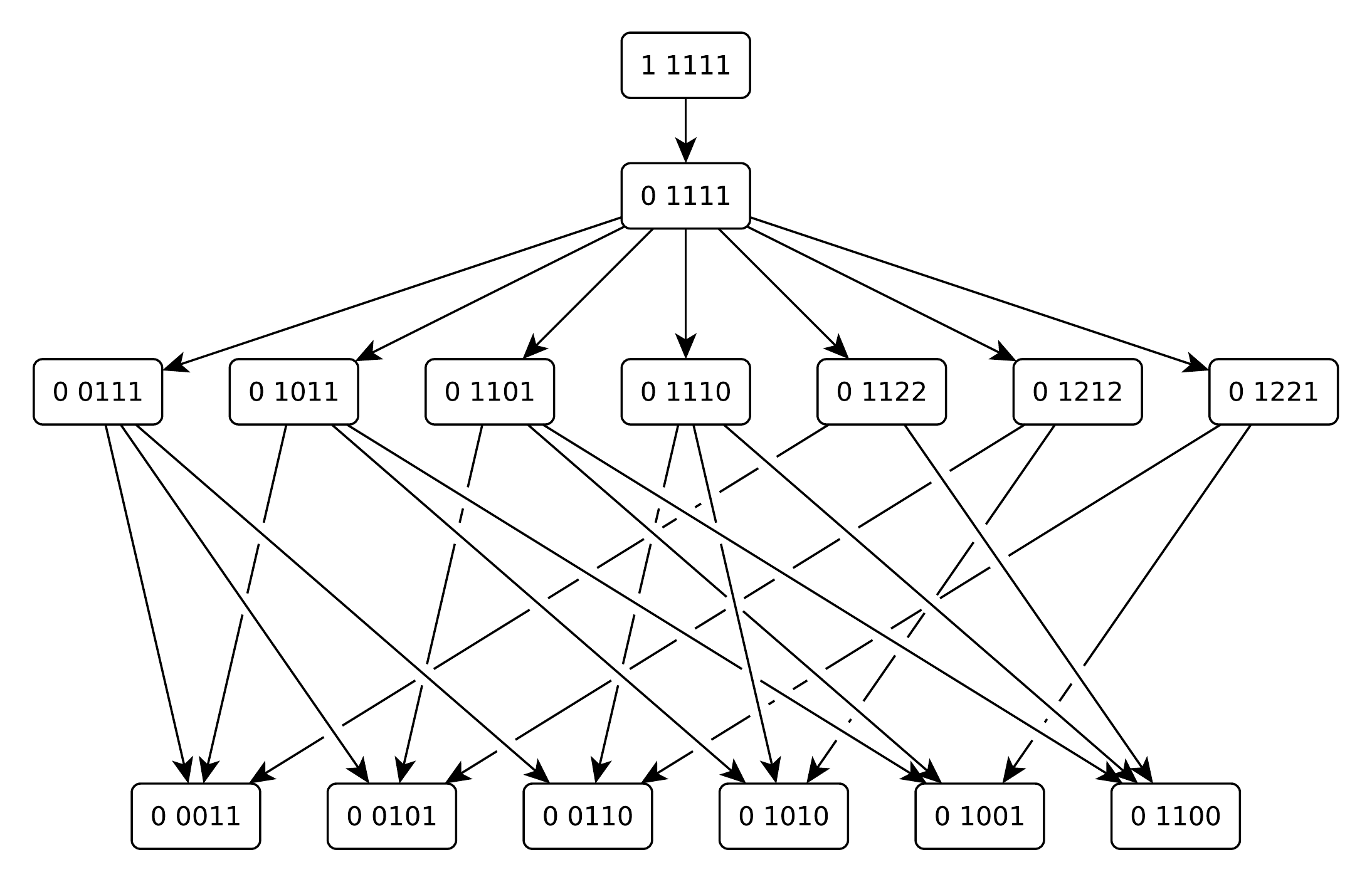}
  \end{center}
  \caption{\label{fig:inmf}Embedding structure of invariant manifolds
    for a starlike network of $N=5$ nodes.}
\end{figure}

The synchronization manifolds have different dimensions and belong to
each other forming a hierarchy. There is a single full synchronization
manifold encoded with a sequence of $N$ ones $\{111\ldots1\}$. Its
dimension is equal to the local dimension of a node oscillator $M$.
This manifold is embraced by $2M$ dimensional manifold of full
synchronization of all subordinates without the hub. The sequence
denoting it has zero on the first position and $N-1$ ones:
$\{011\ldots1\}$. Then there is a set of $3M$ dimensional manifolds
containing clusters where all but one of the subordinates are
synchronized. The encoding sequences are: $\{0011\ldots1\}$,
$\{0101\ldots1\}$, $\{0110\ldots1\}$, and so on. There are $N-1$
manifolds of this type. If $N\geq 5$, there are $3M$ dimensional
manifolds corresponding to two clusters. These are $\{02211\ldots1\}$,
$\{02121\ldots1\}$, $\{02112\ldots1\}$, and so an. The number of such
manifolds is $(N-1)!/[i!(N-1-i)!]$, where $2\leq i\leq N-3$ is the
size of the second cluster. The manifold $\{011\ldots1\}$ is the
intersection of all $3M$ dimensional manifolds. The depth of this
hierarchy depends on $N$. On the bottom level there are manifolds of
dimension $(N-1)M$ containing clusters of two synchronized
subordinates. The encoding sequences are $\{01100\ldots 0\}$,
$\{00110\ldots 0\}$, $\{00011\ldots 0\}$, and so
on. Figure~\ref{fig:inmf} illustrates this showing the complete hierarchy
for $N=5$.

\section{Global master stability function for full
  synchronization}\label{sec:glob_msf}

The stability of the synchronization manifold
$\vec x_1=\cdots=\vec x_N=\vec x$ can be analyzed using so called
master stability function (MSF)~\cite{MSF98,Boccaletti2006175}. We
have to remind however, that for chaotic dynamics this analysis
provides only a necessary condition. Due to the embedded into
synchronous chaotic attractor limiting sets whose transverse stability
can differ from the average stability of the attractor as a whole, the
destruction of the full synchronization can occur even when MSF
indicates the stability~\cite{milnor2004concept,Ott199439}.

At the synchronization manifold elements of $\mat B_{mi}$ are reduced
to diagonal matrices with identical elements:
\begin{equation}
  \mat F_{mi}=f^{(m)}_{i}\identmat,\;\;
  \mat \Phi_{m}=f^{(m)}_{M+1}\identmat,\;\;
  \mat G_{mi}=g^{(m)}_{i}\identmat,
\end{equation}
where
\begin{equation}
  f^{(m)}_{i}\equiv \frac{\partial}{\partial x^{(i)}}f^{(m)}(\vec x,0),\;\;
  f^{(m)}_{M+1}\equiv \frac{\partial}{\partial h}f^{(m)}(\vec x,0),\;\;
  g^{(m)}_{i}\equiv \frac{\partial}{\partial x^{(i)}}g^{(m)}(\vec x).
\end{equation}
Let $\mat U$ be a column matrix of eigenvectors of $\mat
L$. Decomposing perturbation vectors $\vec{\delta x}^{(m)}$ over the
eigenvectors $\mat U$ and taking into account that on the
synchronization manifold all matrices $\mat B_{mi}$ are simultaneously
diagonalized by $\mat U$, one obtains an $M$ dimensional map for
perturbations to synchronous attractor:
\begin{equation}
  \label{eq:msf_gen}
  \delta y^{(m)}(t+1)=\sum_{i=1}^M\mu_{mi}(\theta)\delta y^{(i)}(t),
\end{equation}
where $\theta$ is an eigenvalue of $\mat L$ and
\begin{equation}
  \mu_{mi}(\theta)=f_i^{(m)}+\epsilon_mf_{M+1}^{(m)}g_i^{(m)}\theta.
\end{equation}

The existence of the full synchronization solution means that at least
one of the eigenvalues of $\mat L$ is zero and the corresponding
eigenvector contains identical elements. This eigenvector is
responsible for longitudinal perturbations to synchronous
attractor. All nonzero eigenvalues correspond to transverse
perturbations and thus are responsible for average stability of full
synchronization.

Thus iterating a single local map and computing the first Lyapunov
exponents using Eq.~\eqref{eq:msf_gen} one obtains global MSF for the
network as $\max\{\lambda_{\text{gmsf}}(\theta)\,|\,\theta>0\}$, i.e.,
the synchronization manifold is stable on average if
$\lambda_{\text{gmsf}}<0$ for any nonzero $\theta$.

\section{Master stability function for cluster synchronization}

Let a synchronized cluster includes subordinate nodes with indexes
$j_1$, $j_2$,\ldots,$j_C$, where $j_n\geq 2$, and $C\leq N-1$ is the
cluster size. It means that $\vec x_{j_1}=\vec x_{j_2}=\cdots=\vec
x_{j_C}$.

The Jacobian matrix is built of cells $\mat B_{mi}$ which, in turn,
are constructed of diagonal matrices $\mat F_{mi}$, $\mat\Phi_{mi}$
and $\mat G_{mi}$, see Eq.~\eqref{eq:jac_f_ph_g} and
\eqref{eq:star_cell}.  When the cluster emerges, the corresponding
diagonal elements $j_1$, $j_2$,\ldots,$j_C$ of these matrices
coincides. This results in the coincidence of elements of the first
column and the diagonal elements of $\mat B_{mi}$ with these
indexes. The first row of $\mat B_{mi}$ will contain identical
elements multiplied by the corresponding weights $w_{j_n}$.  Let, for
example, $N=5$ and the cluster include three subordinates, $C=3$:
$j_1=2$, $j_2=3$, and $j_3=4$. Then the cell $\mat B_{mi}$ of the
Jacobian matrix has the following structure:
\begin{equation}
  \mat B_{mi}=
  \left(
    \begin{array}{lllll}
      b_{mi11} & w_2 q_{mi} & w_3 q_{mi} & w_4 q_{mi} & b_{mi15} \\
      r_{mi}   & p_{mi}    & 0         & 0         & 0 \\
      r_{mi}   & 0        & p_{mi}     & 0         & 0 \\
      r_{mi}   & 0        & 0         & p_{mi}     & 0 \\
      b_{mi51} & 0        & 0          & 0        & b_{mi55}
    \end{array}
  \right)
\end{equation}

To verify stability of the cluster we have to consider the evolution
of transverse tangent perturbations to its trajectories.  Consider a
perturbation vector represented in block form as $\vec{\delta
  x}_\transv= \left(\vec{\delta x}^{(1)},\vec{\delta x}^{(2)},\ldots,
  \vec{\delta x}^{(M)}\right)^\transp$.  Each element of this vector
has zeros everywhere except for the sites $j_1$, $j_2$,\ldots,$j_C$
corresponding to the cluster:
\begin{equation}
  \label{eq:perp_pert}
  \vec{\delta x}^{(i)}=\left(\ldots,0,
    \delta x_{j_1}^{(i)},
    \delta x_{j_2}^{(i)},
    \ldots,
    \delta x_{j_C}^{(i)},0,\ldots\right)
\end{equation}
Moreover, the following holds:
\begin{equation}
  \label{eq:perp_pert_sum}
  \sum_{n=1}^Cw_{j_n}\delta x_{j_n}^{(i)}=0.
\end{equation}
One can easily check that $\vec{\delta x}^{(i)}$ given by
Eqs.~\eqref{eq:perp_pert}, and~\eqref{eq:perp_pert_sum} is an
eigenvector of $\mat B_{mi}$ with the eigenvalue $p_{mi}$:
\begin{equation}
  \mat B_{mi}\vec{\delta x}^{(i)}=\vec{\delta x}^{(i)}p_{mi}.
\end{equation}
In view of these properties the evolution of the tangent vectors
$\vec{\delta x}_\transv$ is described as follows:
\begin{equation}
\label{eq:perp_jacobian_step}
    \vec{\delta x}^{(m)}(t+1)=\sum_{i=1}^M\mat B_{mi}\,\vec{\delta
      x}^{(i)}(t)=\sum_{i=1}^M p_{mi}\vec{\delta x}^{(i)}(t).
\end{equation}
Thus, the Jacobian matrix $\mat J$ acts on transverse perturbation
vector $\vec{\delta x}_\transv(t)$ in the same ways as a block matrix
$\mat J'$ build of $M\times M$ diagonal cells $\mat B'_{mi}$ with
identical diagonal elements:
\begin{equation}
  \mat B'_{mi}=p_{mi}\identmat.
\end{equation}

To conclude if the cluster is stable, we need to compute Lyapunov
exponent corresponding to the perturbation vector $\vec{\delta
  x}_\transv(t)$. In course of the computations we first normalize
$\vec{\delta x}_\transv(t)$, make a step with the Jacobian matrix as
defined by Eq.~\eqref{eq:perp_jacobian_step} and find a norm of the
resulting vector $\vec{\delta x}^{(m)}(t+1)$ which is
\begin{equation}
  \left\|\vec{\delta x}_\transv(t+1)\right\|^2=
  \sum_{i,j,m} p_{mi}p_{mj}
  \left[\vec{\delta x}^{(j)}(t)\right]^\transp\vec{\delta x}^{(i)}(t).
\end{equation}
Lyapunov exponent is an average logarithms of these norms. Important
is that the norm does not depend on the number of nodes $N$. Moreover,
the theory of Lyapunov exponents establishes that they do not depend
on the choice of the initial vector $\vec{\delta
  x}_\transv(0)$~\cite{CLV2012}. Thus, to check the transverse
stability of the cluster, it is enough to iterate matrices $\mat
P=\{p_{mi}\}$ with non-block tangent vector of $M$ elements
periodically performing its re-normalization,
\begin{equation}
\label{eq:gmsf_jacobian_step}
    \delta x^{(m)}(t+1)=\sum_{i=1}^M p_{mi}\delta x^{(i)}(t).
\end{equation}
We will refer the Lyapunov exponent, computed as average logarithm of
norms of this vector as a cluster master stability function (CMSF).
Elements of $\mat P$ are taken from the matrix~\eqref{eq:star_cell}
where elements are defined by Eqs.~\eqref{eq:jac_f_ph_g}.
\begin{equation}
  \label{eq:pmi}
  p_{mi}=\frac{\partial f^{(m)}}{\partial x^{(i)}}(\vec x,h^{(m)})-
  \epsilon_m \frac{\partial f^{(m)}}{\partial h}(\vec x,h^{(m)})
  \frac{\partial g^{(m)}}{\partial x^{(i)}}(\vec x).
\end{equation}
Here $\vec x$ and $h^{(m)}$ is computed at the corresponding
synchronization cluster.

\section{Cluster related network reduction}

Since any cluster belongs to an invariant manifold, theoretically,
initially identical variables corresponding to the cluster have to be
identical forever, i.e.,
$\vec x_{j_1}(t)=\vec x_{j_2}(t)=\cdots=\vec x_{j_C}(t)$ for any
$t$. It is well known, however, that in actual computations this may
not be the case. When the synchronization cluster is transversely
unstable an unavoidable numerical noise due to round-off errors
destroys it. Nevertheless the deviation from the invariant manifold
afthe er one time step is small. Thus, to compute $\vec x$ and
$h^{(m)}$ for CMSF of a given cluster one can correct this deviation
at each step by assigning to cluster variables identical values equal
to their average.

Being the most simple and straightforward this method is not the
optimal since requires redundant computations. One can instead iterate
a reduced network where a set of cluster nodes with identical states
is represented by a single node. The weight of this node link is equal
to the sum of weights of initial nodes belonging to the cluster.

\begin{figure}
  \begin{center}
    \widfig{0.5}{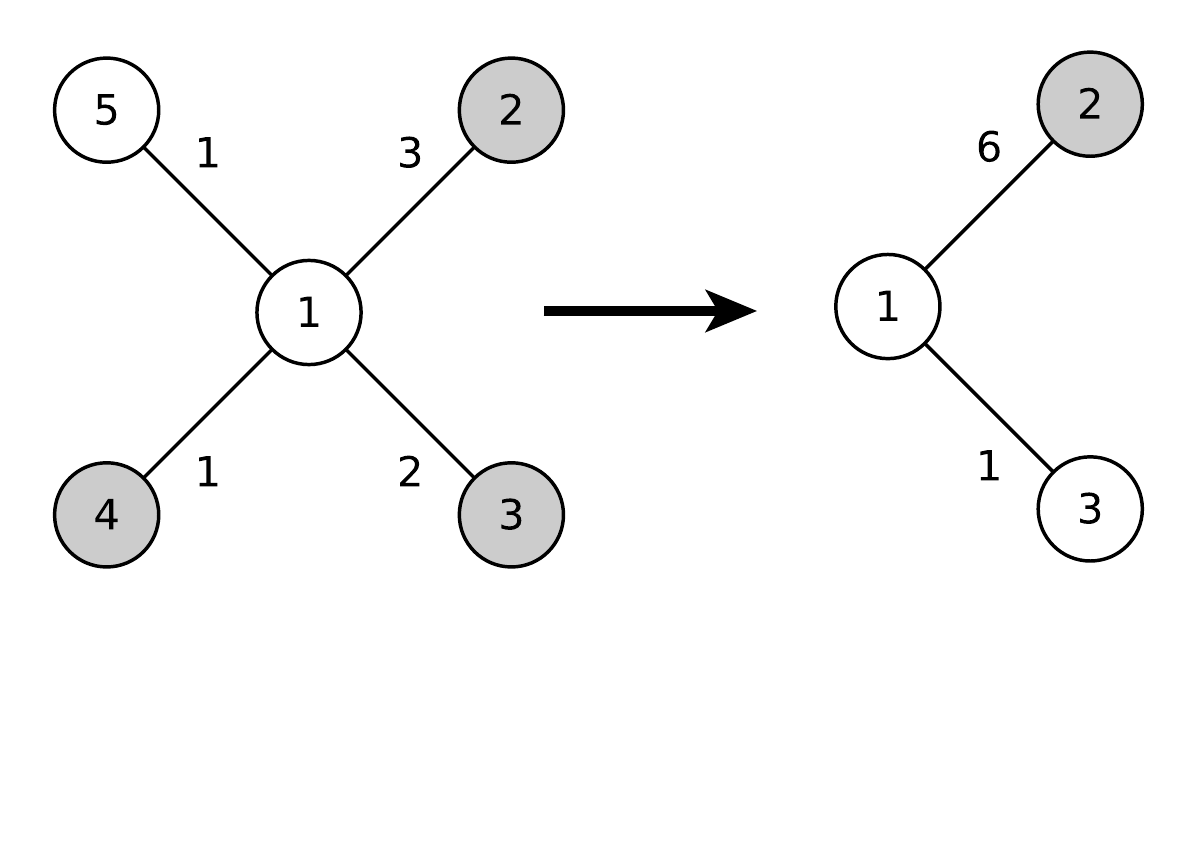}
  \end{center}
  \caption{\label{fig:starred}Reduction of a starlike network. If the
    nodes 2, 3, and 4 get synchronized they can be changed with the
    single node whose link weight is 6.}
\end{figure}

Figure~\ref{fig:starred} illustrates the reduction of the star with
$N=5$ and weights $w_2=3$, $w_3=2$, $w_4=1$, $w_5=1$. If the nodes 2,
3 and 4 get synchronized, the network dynamics will be reproduced by
the reduced network with $N=3$. The node 2 of this new network
represents the cluster having the weight $w'_2=w_2+w_3+w_4$. 

To understand why this reduction is possible, consider $h_1^{(m)}$
for the original network, shown in Fig.~\ref{fig:starred}:
\begin{equation}
  h_1^{(m)}=\epsilon_m\left\{
    \left[
      3g^{(m)}(\vec x_2)+
      2g^{(m)}(\vec x_3)+
      g^{(m)}(\vec x_4)+
      g^{(m)}(\vec x_5)
    \right]/7-g^{(m)}(\vec x_1)
  \right\}.
\end{equation}
When the nodes 2, 3 and 4 are synchronized,
$g^{(m)}(\vec x_2)=g^{(m)}(\vec x_3)=g^{(m)}(\vec x_4)$ and
$h_1^{(m)}$ reads:
\begin{equation}
  h_1^{(m)}=\epsilon_m\left\{
    \left[
      6g^{(m)}(\vec x_2)+
      g^{(m)}(\vec x_5)
    \right]/7-g^{(m)}(\vec x_1)
  \right\}.
\end{equation}
Since $h_n^{(m)}$ with $n\geq 2$ are given by
Eq.~\eqref{eq:netw_h_star2} regardless of $N$ and $w_j$, one can
reproduce the dynamics in presence of the cluster by substituting the
three original nodes with the one whose link has weight 6.

In some cases one more reduction step can be made. Consider, for
example, the network of $N=5$ nodes with weights
$w_2=w_3=w_4=w_5=1$. Let there are two clusters such that subordinate
nodes get synchronized pairwise: 2nd with 3rd and 4th with 5th. This
network is reduced to the network with $N=3$ nodes whose link have
weights $w_2=w_3=2$. Now since the coupling is normalized so that each
$h_n^{(m)}$ with $n\geq 2$ is always given by
Eq.~\eqref{eq:netw_h_star2}, the original synchronous dynamics will be
reproduced by the network with $N=3$ and weights $w_2=w_3=1$.

This can be treated in reciprocal order: Each starlike network with
integer positive weights correspond to a larger network some of whose
nodes are synchronized.

\section{Examples}

\subsection{Ikeda starlike network}\label{sec:netw_iked}

This map was suggested by Ikeda\cite{Ikeda1} and Ikeda et
al.\cite{Ikeda2}. We build a starlike network of Ikeda maps and
introduce the coupling between the nodes of Ikeda network using ideas
reported by Otsuka and Ikeda\cite{IkedaCoupling1} and by
Otsuka~et~al.~\cite{IkedaCoupling2}:
\begin{equation}
  \label{eq:ikeda}
  z_n(t+1)=\alpha+\beta w[z_n(t)]+\epsilon
  \left(
    \sum_{j=1}^N\frac{a_{nj}}{k_n}w[z_j(t)]-w[z_n(t)]
  \right),
\end{equation}
where $z_n(t)$ is a complex variable and
$w(z)=z\mee^{\mii(|z|^2+\phi)}$. In what follows the parameters of
this system will be $\alpha=2$, $\beta=0.5$, $\phi=0.3$. For these
values a single uncoupled system demonstrates chaos with the Lyapunov
exponents $\lambda_1=0.868$, $\lambda_2=-2.254$.

As discussed by Kuptsov and Kupstova\cite{WildStars}, the phase space
of a starlike network can contain spuriously stable limit sets. They
emerge due to round-off errors in computations. To eliminate them, a
very small noise of the amplitude $10^{-12}$ will be added to
variables at each step. This noise is found to be too small to result
in any observable changes of the dynamics, however it is enough to
eliminate spurious regimes.

For the Ikeda network the generalized
model~\eqref{eq:netw_gen},~\eqref{eq:netw_gen_h} takes a form:
$f^{(1)}(\vec x,h)=\alpha+\beta u(x^{(1)},x^{(2)})+h$,
$f^{(2)}(\vec x,h)=\beta v(x^{(1)},x^{(2)})+h$,
$g^{(1)}(\vec x)=u(x^{(1)},x^{(2)})$,
$g^{(2)}(\vec x)=v(x^{(1)},x^{(2)})$, where
$u(x^{(1)},x^{(2)})=x^{(1)}\cos\theta-x^{(2)}\sin\theta$,
$v(x^{(1)},x^{(2)})=x^{(1)}\sin\theta+x^{(2)}\cos\theta$,
$\theta=(x^{(1)})^2+(x^{(2)})^2+\phi$,
$\epsilon_1=\epsilon_2=\epsilon$.

According to the results reported by Kuptsov and
Kupstova\cite{WildStars}, starlike networks can demonstrate very rich
multistability. To distinguish various regimes emerging and vanishing
as the coupling strength changes, we will use the first Lyapunov
exponent. Being positive, it indicates the presence of chaos; the
negative sign reveals periodicity; and the zero signals
quasi-periodicity. It is very unlikely (though, of course, not totally
excluded) that different regimes will have identical Lyapunov
exponents. Thus, to reveal a sort of dynamics we will take a pool of
random initial conditions and compute $\lambda_1$ for each
corresponding trajectory. Grouping of the resulting values near a few
point indicates the presence of multiple regimes. This approach is
often used for analysis of
multistability~\cite{Pisarchik2014167,WildStars}.

\begin{figure}
  \begin{center}
    \widfig{0.6}{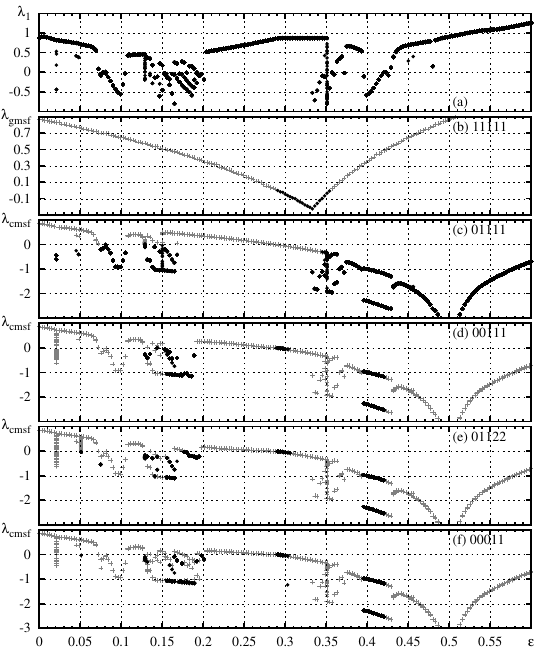}
  \end{center}
  \caption{\label{fig:gmsf_ikeda}Regimes of Ikeda starlike
    network~\eqref{eq:ikeda} and their stability: the first Lyapunov
    exponent $\lambda_1$ (a), global MSF (b) and CMSF (c,d,e,f) for
    all possible clusters admitted at $N=5$.}
\end{figure}

Figure~\ref{fig:gmsf_ikeda}(a) shows the first Lyapunov exponent of
the Ikeda network with $N=5$ and unit weights of links computed for
various initial conditions vs. $\epsilon$. One can observe that
chaotic regimes dominate, but also there are areas where dynamics is
periodic. Many different $\lambda_1$ at the same $\epsilon$ reveals
multistability.

Figure~\ref{fig:gmsf_ikeda}(b) shows the global MSF computed as
described in Sec~\ref{sec:glob_msf}. Solid bullet points mark negative
$\lambda_{\text{gmsf}}$, while gray crosses represent positive
ones. The area of stability is
\begin{equation}
  \label{eq:gmsf_area_ikeda}
  0.288<\epsilon<0.357.
\end{equation}
These boundaries are determined from the straightforward inspection of
the data computed for Fig.~\ref{fig:gmsf_ikeda}(b). The precision is
$0.003$. Obviously, one can compute them more precisely, however this
sufficiently low precision is enough for us.

Figures~\ref{fig:gmsf_ikeda}(c,d,e,f) represent CMSFs for all clusters
admitted at $N=5$. The figure legends contains encoding sequences
labeling corresponding invariant manifold in the same way as in
Fig.~\ref{fig:inmf}. Black bullet point is plotted when the
corresponding cluster is stable, $\lambda_{\text{cmsf}}<0$, and other
types of synchronization are absent, i.e., we are outside of the
embracing synchronization manifold, see the discussion in
Sec.~\ref{sec:manif}. Otherwise gray crosses are plotted.

Within the area given by inequalities~\eqref{eq:gmsf_area_ikeda} the
full synchronization regime is expected to be observed, since it is
transversally stable on average. However, as we already mentioned
above, this is the case only for regular dynamics, while for full
chaotic synchronization global MSF is necessary but not sufficient
condition. 

\begin{figure}
  \begin{center}
    \widfig{0.6}{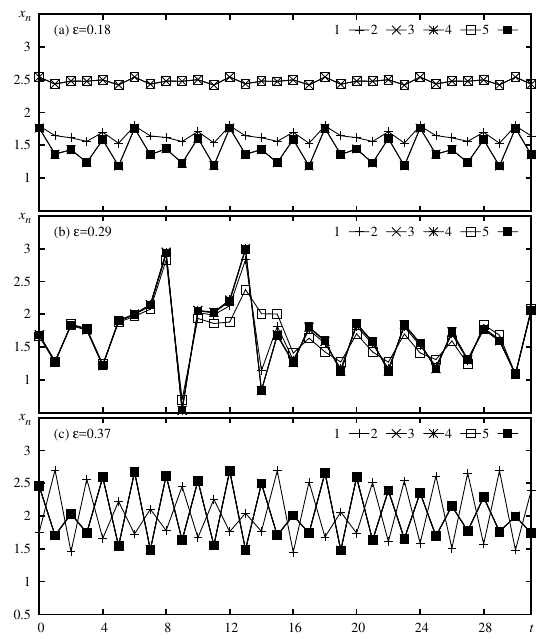}
  \end{center}
  \caption{\label{fig:timser_ikeda}Time series of Ikeda starlike
    network~\eqref{eq:ikeda}.}
\end{figure}

Comparing Figs.~\ref{fig:gmsf_ikeda}(a) and (b) we observe that the
left boundary of the stability of the global synchronization regime in
Fig.~\ref{fig:gmsf_ikeda}(a) very well coincides with the predicted
one via MSF. (One can easily distinguishes the regime of the full
chaotic synchronization in Fig.~\ref{fig:gmsf_ikeda}(a) as a
horizontal line at level of $\lambda_1$ coinciding with $\lambda_1$
for uncoupled maps at $\epsilon=0$.) However at the right edge the
full synchronization disappears earlier. First, one can see in
Fig.~\ref{fig:gmsf_ikeda}(b) that the cluster $\{01111\}$ becomes
stable already at $\epsilon=0.33$. This results in the multistability
when this cluster coexists with the full synchronization. An
illustration of this regime is shown in
Fig.~\ref{fig:timser_ikeda}(c). At $\epsilon=0.35$ Lyapunov exponents
in Fig.~\ref{fig:gmsf_ikeda}(a) are arranged along a straight line
ranging approximately from $-1$ to $1$. The most reasonable
explanation of this volatility is the presence of called unstable
dimension variability (UDV)~\cite{UDV}. It is known that typically on
the edges of stability of full synchronization regimes there are a lot
of periodic orbits with different dimensions of unstable manifold. The
trajectory passes in vicinities of these orbits changing its own
unstable manifold dimension. In particular it results in very bad
convergence of Lyapunov exponents.

Moving further to the right outside of the
range~\eqref{eq:gmsf_area_ikeda} we observe in
Fig.~\ref{fig:gmsf_ikeda}(c) the presence of multistability of regimes
with fully synchronized subordinate nodes. They can be both chaotic
and regular.

The area $0.39<\epsilon 0.42$ is highlighted by black bullet points
simultaneously in all Figs~\ref{fig:gmsf_ikeda}(c,d,e,f). This is a
specific reaction of the computation algorithm to so called
oscillation death that takes place here. All nodes do not oscillate,
the subordinate stay in one point and the hub has another state.
Notice two values of $\lambda_{\text{cmsf}}$ here. They correspond to
the interchange of the states of the hub and the subordinates.

Further to the right the cluster with synchronized subordinate nodes
remain the only stable regime, see black bullet points in
Fig.~\ref{fig:gmsf_ikeda}(c) at $\epsilon>0.42$.

On the left edge of the synchronization area, at approximately
$0.29<\epsilon\approx 0.3$ Figs.~\ref{fig:gmsf_ikeda}(d,e,f) indicate
simultaneous stability of clusters with three or two nodes. However no
multistability is registered in Fig.~\ref{fig:gmsf_ikeda}(a). More
detailed inspection reveals that here is an area of
intermittency. Nodes get almost synchronized for some time, but then a
desynchronization occurs. An example of this behavior is shown in
Fig.~\ref{fig:timser_ikeda}(b).

To the left of the full synchronization
area~\eqref{eq:gmsf_area_ikeda} we observe first a chaotic regime
without multistability, see Fig.~\ref{fig:gmsf_ikeda}(a). The gray
crosses on on the panels below, Fig.~\ref{fig:gmsf_ikeda}(b,c,d,e,f)
indicate that no synchronization clusters can happen here.

Below $\epsilon=0.2$ there is an area of
multistability. Fig.~\ref{fig:gmsf_ikeda}(a) shows here a variety of
regimes, both chaotic and regular. Black bullet points in
Figs.~\ref{fig:gmsf_ikeda}(d,e,f) indicate that this multistability is
related with the presence of synchronization clusters of different
types. The emergence of the rich multistability area below the full
synchronization regime was already reported by Kuptsov and Kuptsova
for a starlike network of H\'enon maps\cite{WildStars}. An example of
dynamics in this area is shown in Fig.~\ref{fig:timser_ikeda}(a).

\subsection{Predator-prey  starlike network}

The predator-prey network can be built analogously to a lattice of
predator-prey maps considered by Sol{\'e} and Valls\cite{PreyPredCML}:
\begin{equation}
  \label{eq:preypred}
  \begin{aligned}
    x_n(t+1)&=\alpha x_n(t)[1-x_n(t)-y_n(t)]+
    \epsilon\left(
      \sum_{j=1}^N\frac{a_{nj}}{k_n}x_j(t)-x_n(t)
    \right),\\
    y_n(t+1)&=\beta x_n(t)y_n(t)+
    d\epsilon\left(
      \sum_{j=1}^N\frac{a_{nj}}{k_n}y_j(t)-y_n(t)
    \right).
  \end{aligned}
\end{equation}
Again a very small noise of the amplitude $10^{-12}$ is added at each
step to destroy spurious regimes. The generalized
model~\eqref{eq:netw_gen},~\eqref{eq:netw_gen_h} for this network
reads: $f^{(1)}(\vec x,h)=\alpha x^{(1)}(1-x^{(1)}-x^{(2)}+h$,
$f^{(2)}(\vec x,h)=\beta x^{(1)} x^{(2)}+h$,
$g^{(1)}(\vec x)=x^{(1)}$, $g^{(2)}(\vec x)=x^{(2)}$, and
$\epsilon_1=\epsilon$, $\epsilon_2=d\epsilon$.

\begin{figure}
  \begin{center}
    \widfig{0.6}{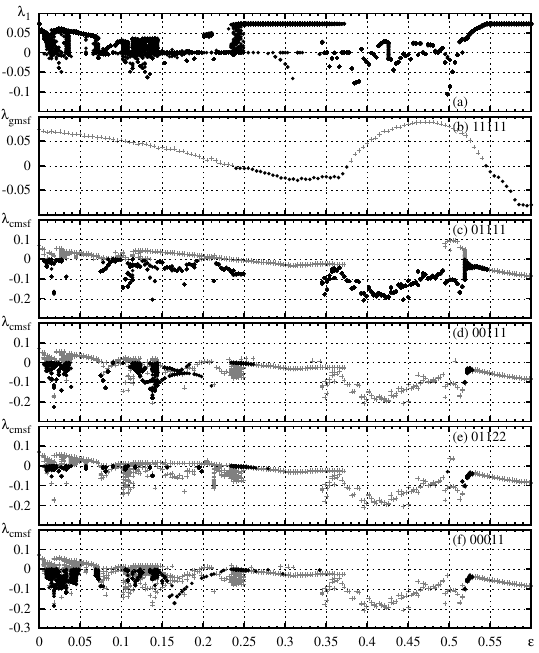}
  \end{center}
  \caption{\label{fig:gmsf_preypred}Same as Fig.~\ref{fig:gmsf_ikeda}
    for predator-prey starlike network~\eqref{eq:preypred}, $N=5$.}
\end{figure}

The first Lyapunov exponent as well as master stability functions are
shown in Fig.~\ref{fig:gmsf_preypred}. First of all notice that there
are two areas where the whole network can be stable, see
Fig.~\ref{fig:gmsf_preypred}(a). Their boundaries are
\begin{equation}
  \label{eq:gmsf_area_preypred}
  0.235<\epsilon<0.380,\;\;
  0.540<\epsilon.
\end{equation}
The oscillations of the network diverges at $\epsilon>0.6$.

\begin{figure}
  \begin{center}
    \widfig{0.6}{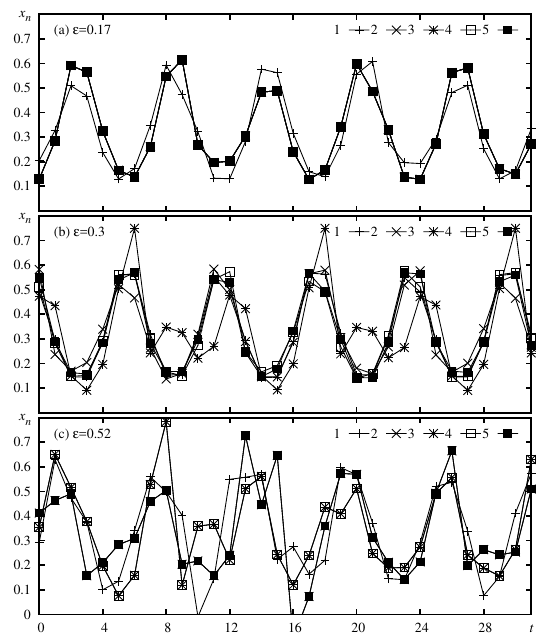}
  \end{center}
  \caption{\label{fig:timser_praypred}Time series of predator-prey
    starlike network~\eqref{eq:preypred}.}
\end{figure}

The multistability is observed almost everywhere, except for the right
full synchronization area. Within the left area there is the second
attractor which is quasiperiodic, since its $\lambda_1$ vanishes. Also
at $\epsilon>0.28$ more attractors coexist with the full
synchronization, both chaotic ($\lambda_1>0$) and periodic
($\lambda_1<0$). An example of chaotic dynamics here is shown in
Fig.~\ref{fig:timser_praypred}(b).

To the right of the first full synchronization area the cluster
$\{01111\}$ becomes stable as above for the Ikeda network. Near the
second area we again observe multistability, see points around
$\epsilon=0.52$ in Figs.~\ref{fig:gmsf_preypred}(c,d,e,f). All sorts
of clusters can be stable here and the inspection of time series
revels that corresponding regimes can indeed be observed. For example,
Fig.~\ref{fig:timser_praypred}(c) shows chaotic oscillations with
synchronized 2nd and 5th nodes as well as 3rd and 4th ones.

To the left of the first full synchronization area we again observe
rich multistability. Figures~\ref{fig:gmsf_preypred}(c,d,e,f)
demonstrates that all sorts of clusters can emerge here. An
illustration of the dynamics is shown in
Fig.~\ref{fig:timser_praypred}(a).

The specific feature of the discussed network is high volatility of
the first Lyapunov exponent at $\epsilon<0.25$, see
Fig.~\ref{fig:gmsf_preypred}(a). Instead of grouping around several
points almost continues ranges of values are observed. We address it
to the rich multistability and UDP related to it. Phase space is
filled up with interwoven limiting sets with different numbers of
unstable manifolds. Since a trajectory passes around these sets its
unstable manifold dimension fluctuates so that the convergence of the
first Lyapunov exponent becomes very bad.

\subsection{H\'enon starlike network}

Starlike network of H\'enon maps can be built as generalization of the
H\'{e}non chain suggested by Politi and Torcini\cite{PolTor92a}:
\begin{equation}
  \label{eq:netw_henon}
  \begin{gathered}
    x_n(t+1)=\alpha-\left[x_n(t)+
    \epsilon \left(
      \sum_{j=1}^N \frac{a_{nj}}{k_n}x_j(t)-x_n(t)
    \right)
    \right]^2+y_n(t),\\
    y_n(t+1)=\beta x_n(t),
  \end{gathered}
\end{equation}
Here $\alpha=1.4$ and $\beta=0.3$ are the parameters controlling local
dynamics. Recall that the H\'{e}non map is time-reversible. The
coupling is introduced in a way that preserves this property. As for
previous networks we add very small noise with the amplitude
$10^{-12}$ to destroy spurious regimes. The general
form~\eqref{eq:netw_gen},~\eqref{eq:netw_gen_h} for this map reads:
$f^{(1)}(\vec x,h)=\alpha-(x^{(1)}+h)^2+x^{(2)}$,
$f^{(2)}(\vec x,h)=\beta x^{(1)}$, $g^{(1)}(\vec x)=x^{(1)}$,
$g^{(2)}(\vec x)=0$, and $\epsilon_1=\epsilon$, $\epsilon_2=0$.

\begin{figure}
  \begin{center}
    \widfig{0.6}{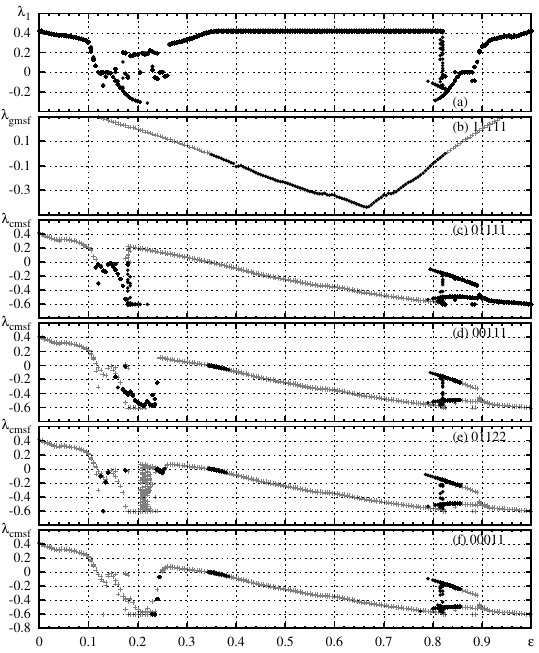}
  \end{center}
  \caption{\label{fig:gmsf_henon}Same as Fig.~\ref{fig:gmsf_ikeda} for
    H\'enon starlike network~\eqref{eq:netw_henon}, $N=5$.}
\end{figure}

We already considered starlike network of H\'enon maps in
detail\cite{WildStars}. In particular time series of all regimes were
presented and discussed. Thus, in the present paper we show only the
first Lyapunov exponent, Fig.~\ref{fig:gmsf_henon}(a), compared with
both the global MSF and CMSFs, Fig.~\ref{fig:gmsf_henon}(b) and
Fig.~\ref{fig:gmsf_henon}(c,d,e,f), respectively.

As for two previous networks we observe the area where the full
synchronization attractor can be stable. It lays within the range
\begin{equation}
  \label{eq:gmsf_area_henon}
  0.348<\epsilon<0.826.
\end{equation}
Notice that the overall picture of dynamics is qualitatively similar
to that for Ikeda map discussed in Sec.~\ref{sec:netw_iked}.  To the
right of the area of the full synchronization there is an area of
oscillation death, $0.8<\epsilon<0.83$. It manifests itself as black
bullet points emerging simultaneously on panels (c), (d), (e) and
(f). More to the right the cluster $\{01111\}$ becomes stable, see
Fig.~\ref{fig:gmsf_henon}(c). To the left of the area of full
synchronization we have an intermittency, see bars of black bullet
points in Figs.~\ref{fig:gmsf_henon}(d,e,f) at
$0.34<\epsilon0.38$. Then to the left of this regime we observe
chaotic oscillations at $0.26<\epsilon<0.34$. And further to the left
we encounter multistability. Here all clusters can be stable, but the
cluster of all subordinates dominates.

\section{Conclusion}

We considered a generalized model of a starlike network with discrete
time oscillators and normalized Laplacian coupling. The coupling
admits both full synchronization and cluster synchronization of
subordinate nodes. We suggested a necessary condition of stability of
clusters. This approach generalizes the well known method of master
stability function developed for analysis of full synchronization.

Three networks are considered and the stability of clusters is
analyzed. The common feature specific to these three systems is that
the area of full synchronization on the axis of the coupling strengths
is surrounded by areas where clusters of synchronized subordinates are
stable. Within these area very rich multistability is observed. 

\acknowledgments

This work was partially supported (P.V.K.) by RF President program for
Leading Russian research schools NSh-1726.2014.2.

\bibliography{constel}
\bibliographystyle{spiebib} 

\end{document}